# Global Stability Analysis for an Internet Congestion Control Model with a Time-Varying Link Capacity

B. Rezaie, M.-R. Jahed Motlagh, M. Analoui
Iran University of Science and Technology
{brezaie,jahedmr,analoui}@iust.ac.ir

S. Khorsandi
Amirkabir University of Technology
khorsand@cic.aut.ac.ir

***Abstract*—In this paper, a global stability analysis is given for a rate-based congestion control system modeled by a nonlinear delayed differential equation. The model determines the dynamics of a single-source single-link network, with a time-varying capacity of link and a fixed communication delay. We obtain a sufficient delay-independent conditions on system parameters under which global asymptotic stability of the system is guarantied. The proof is based on an extension of Lyapunov-Krasovskii theorem for a class of nonlinear time-delay systems. The numerical simulations for a typical scenario justify the theoretical results.**

## I. Introduction

In recent years, the Internet grows explosively, and there has been a critical necessity for efficient schemes to overcome the congestion problem as the number of network users continues to increase. The Internet congestion control system employs rate controllers at the sources and the routers, which are often modeled by Nonlinear Delayed Differential Equations (NDDEs) [1]. The existence of nonlinearity and delay in the congestion control system leads to undesirable behavior and instability that can degrade the network performance [2]. Therefore, there has been a surge of interest to improve the congestion control mechanisms in order to ensure the global stability and the performance of the network. However, the global stability analysis of NDDEs is a difficult problem. Therefore, researchers are constrained to the investigation of the properties of their nonlinear undelayed versions, or the linearized delayed ones [3-10]. In this paper, we present a novel approach for global stability analysis of the network in the presence of nonlinearity, delay and a time-varying parameter.

The network stability analysis has been extensively studied since the work of Kelly *et al*. [3]. It has been proven that the congestion control algorithms are globally asymptotically stable in the absence of delays [3-6]. In [3,4] a stability condition was derived for a single proportionally fair congestion controller with delayed feedback. Since then, the linear stability properties of congestion control algorithms have been investigated in the presence of non-negligible delays [7-10]. The stability conditions studied in these works are based on linearized system and local stability analysis. Analysis based on linearization is usually misleading because the effect of nonlinearity is ignored.

Global stability analysis of NDDE models in the presence of delays and nonlinearities is difficult. However, this problem has been addressed in several papers [11-25]. Most results concentrate on the Lyapunov-Razumikhin (L-R) stability theory, since the construction of Lyapunov-Krasovskii (L-K) functional is more difficult than that of L-R function [26]. The Lyapunov functions in L-R based methods can only account for a small amount of the energy in the system and they can not account for the energy hidden by the delay.

In this paper, we analyze Kelly's optimization framework [3] for a rate allocation problem and derive the delay independent global stability conditions. Our analysis provides new results for the global stability of networks with delay and time-varying link capacity. The approach is based on an extension of L-K method of global stability analysis for networks as a system described by NDDEs with a fixed delay and a time-varying parameter. Unlike the most of previous works, the obtained conditions are delay-independent and, as a result, the network is stable and robust with arbitrary communication delay. We also consider that the capacity of the network link is time-varying. We assume that current link capacity depends on the rate of the source at current and previous times and the capacity of the link at previous times.

The paper is organized as follows. In Section II, the rate-based model of congestion control system is given. The stability criteria for the congestion control algorithm with delay and time-varying link capacity is obtained in Section III. In Section IV, an example is simulated to show the results for a simple scenario. Finally concluding remarks are given in Section V.

## II. The Model

### *A. Rate-Based Congestion Control Problem*

We consider a fluid model of a network whose topology characterized by a single source/user and a single link/resource. The source has an associated transmission rate $x(t)$. The information from the source passes through a link. The link (wired or wireless) has time-varying capacity $c(t)$.

The source's transmission rate $x(t)$ satisfies the bounds $0 < x_{\min} \le x(t) \le x_{\max}$. The upper bound on the source's flow rate may be a user-specific physical limitation, and the lower bound is due to the fact that the source needs to probe the congestion level of the network by continually transmitting packets. Each source attains a utility of $U(x)$ which is an

This work was supported in part by the Iran Telecommunication Research Center (ITRC).

increasing, strictly concave and continuously differentiable function over the range $x>0$ [3]. As an example, a weighted proportionally fair utility function is of the form $U(.)=w\log(.)$, where $w$ is the weight of the source flow. In addition, the link charges a price per unit described by a function $p(.)$ which is a strictly increasing and continuously differentiable function of the rate going through the link. The price of the link may depend on the link congestion, the loss probability, etc. We also suppose that the transmission of information between the source and the link is delayed due to the link propagation.

The aim of congestion control scheme is to adjust the source's flow rate in order to maximize the utility over the source rate $x(t)$, subject to the constraints of the link capacity, i.e., $x(t) \le c(t)$ . Formally, the rate control problem in the optimization framework proposed in [3] is given by

$$\begin{aligned} &Maximize \quad U(x) - \int_0^x p(y)dy \\ &Subject \quad to \quad x(t) \le c(t) \end{aligned} \tag{1}$$

Kelly *et al.* [3] show that there always exists a transmission rate of source such that uniquely solves the above optimization problem. The unique solution of this optimization problem can be obtained using a gradient projection algorithm [5]. The unique solution of the utility maximization problem (1) is the equilibrium point of primal algorithm.

### B. The Primal Algorithm

In this paper, we consider a model of the primal algorithm for studying global stability. We assume that the price function of the link depends on the link capacity. The capacity of a link in prior works is assumed to be fixed, i.e., in the optimization problem $c$ is not a time-varying function [19,25]. In this paper, we assume that the link has time-varying capacity. Evidently, time-varying capacity demand new investigation on the stability and optimality of traffic congestion control algorithms.

In addition, throughout the paper we assume that source's flow rate $x \in [x_{\min}, x_{\max}]$ and that when the rate of a flow reaches the upper bound $x_{\max}$ (lower bound $x_{\min}$ ), the time derivative of the flow is given by $\dot{x} = \min\{\dot{x}(t), 0\}$ ( $\dot{x} = \max\{\dot{x}(t), 0\}$ ). Therefore, we consider a model of the primal algorithm described by the following delayed differential equation

$$\dot{x}(t) = \kappa\Big[x(t)U'\big(x(t)\big) - x(t)p\big(c(t), x(t)\big)\Big] \tag{2}$$

where $\kappa$ is a positive gain parameter.

The feedback information from the link to the source, which is typically carried by acknowledgments, is delayed due to link propagation. Let $\tau^f$ denotes the forward delay that the source packets endure before reaching the link and $\tau^r$ denotes the reverse delay of the feedback signal from the link to the source. Let us define $\tau = \tau^f + \tau^r$ as the total round-trip delay before the receipt of the acknowledgement of a packet. The price of the link at time $t$ depends on the rate of the source at time $t-\tau^f$ due to the delay from the source to the link, and the rate of the source is adjusted based on the information of the link at time $t-\tau^r$ due to the delay from the link to source. Under these assumptions, the end user dynamical model is given by

$$\dot{x}(t) = \kappa\Big(x(t)U'(x(t)) - x(t-\tau)p(c(t-T), x(t-\tau))\Big) \tag{3}$$

where $T = \tau^r$ .

In this paper, we consider the primal algorithm described by the model (3) and establish out stability analysis.

## III. GLOBAL STABILITY ANALYSIS

We now determine the sufficient delay-independent conditions for global stability of a single link single source network with fixed delay and time-varying link capacity described by (3).

Let the utility function be $U(x(t)) = -1/(ax(t)^a)$ and let the price function be $p(x(t)) = h.\big(x(t)/c(t)\big)^b$ , where $\kappa$ , $a$, $b$ and $h$ are the positive parameters of utility and price functions. Throughout the paper we suppose that $h$=1. In addition, let the capacity of link be state-dependent. Therefore, we get the following equation as a model of a single source/link network

$$\begin{aligned} &\dot{x}(t) = \kappa\left(\frac{1}{x(t)^a} - \frac{x(t-\tau)^{b+1}}{c(t-T)^b}\right) \\ &c(t) = g\big(x(t)\big) \end{aligned} \tag{4}$$

where $g(.)$ can be a nonlinear function. We also consider the following assumptions

- (A1) $\kappa$, $a$, $b$, $\tau$, and $T$ are positive constants, and $\tau \ge T$
- (A2) the network model in (4) has a nonzero and positive initial conditions $x_0(\theta)$ and $c_0(s)$ , which are continuous on $\theta \in [-\tau, 0]$ and $s \in [-T, 0]$.
- (A3) the function $g(.)$ is continuous, positive, and decreasing function, and satisfy the following conditions for all $x$:

$$g(x) > 1,\ g'(x) < -1$$

The equilibrium point of the system satisfies $g(x^*) = x^{*(a+b+1)/b}$ , where $x^*$ is corresponding local equilibrium point of $x(t)$.

For completeness and convenience, we first introduce L-K theory for the case of a nonlinear and unforced time-delay system with a fixed delay. We have the following definitions

Function $\alpha(.) : \Re^+ \to \Re^+$ is said to be of class $\kappa$ if it is continuous, strictly increasing and $\alpha(0)=0$. It said to be of class $\kappa_\infty$ if it is of class $\kappa$ and it is unbounded.

$C$:$C([-\tau,0],\Re)$ is the set of continuous functions mapping the interval $[-\tau,0]$ into $\Re$ with norm $\|x_t\|_\infty = \sup_{\theta\in[-\tau,0]} |x_t(\theta)|$ where $x_t \in C$ .

Let us, consider the following equation

$$\dot{x} = f(x, x_t) \tag{5}$$

where $f$:$\Re\times C\to\Re$.

Suppose $V : C \to \Re^+$ is a continuously differentiable functional and define the right hand derivative of $V$ along the solution of (5) by

$$\dot{V} = \lim_{h\to 0^+}\sup \frac{1}{h}[V(\phi_h^*) - V(\phi)]$$

$$\phi_h^*(s) = \begin{cases} \phi(s+h) & s\in[-\tau,-h] \\ \phi(0) + f(x,x_t)(h+s) & s\in[-h,0] \end{cases}$$

The stability analysis of the system described by (5) has been provided by L-K theorem, which is stated as follows.

*Theorem 1.* (Lyapunov-Krasovskii) (Hale and Verduyn Lunel, 1993) Let $\alpha(.)$ be functions of class $\kappa_\infty$ ; $\beta(.)$ be of class $\kappa$ and $\gamma(.)$ be a non-decreasing function. If there exists a continuously differentiable functional $V$ satisfies

(*i*) $\alpha(|\phi(0)|) \le V(\phi) \le \beta(\|\phi\|_\infty)$ , for all $\phi \in C$ ;

(*ii*) $\dot{V}(\phi) \le -\gamma(|\phi(0)|)$ , for all $\phi \in C$

then, the system (5) is stable. Further, if $\gamma(.)$ is of class $\kappa$ , then the system is asymptotically stable.

Now, we propose the main result of the paper by the following theorem.

*Theorem 2.* Under assumptions (A1)-(A3), if the inequality $\frac{x^{*-a} - x(t)^{-a}}{x(t)-x^*} > \frac{x(t)^{b+1}c(t)^{-b} - x^{*b+1}c^{-b}}{x(t)-x^*}$ holds for each $x$, then the system (4) is globally asymptotically stable.

*Proof.* Define the following L-K functional

$$V = |x(t) - x^*| + \int_{-\tau}^{0} \kappa \frac{x(t+\theta\tau)^{b+1}c(t+\theta T)^{-b} - x^{*b+1}c^{*-b}}{x(t)-x^*} |x(t)-x^*| d\theta$$

It can be easily shown that this L-K functional is continuously differentiable and satisfies the hypothesis (*i*) of Theorem 1. In addition, we have

$$\begin{aligned}\dot{V} &= \dot{x}(t)\frac{|x(t)-x^*|}{x(t)-x^*} \\ &+ \kappa\Bigg[\frac{x(t)^{b+1}c(t)^{-b} - x^{*b+1}c^{*-b}}{x(t)-x^*} \\ &\quad - \frac{x(t-\tau)^{b+1}c(t-T)^{-b} - x^{*b+1}c^{*-b}}{x(t)-x^*}\Bigg]|x(t)-x^*| \\ &= \kappa\Bigg[\frac{x(t)^{-a} - x^{*-a}}{x(t)-x^*} - 2\frac{x(t-\tau)^{b+1}c(t-T)^{-b} - x^{*b+1}c^{*-b}}{x(t)-x^*} \\ &\quad + \frac{x(t)^{b+1}c(t)^{-b} - x^{*b+1}c^{*-b}}{x(t)-x^*}\Bigg]|x(t)-x^*|\end{aligned}$$

Therefore, we obtain

$$\begin{aligned}\dot{V} &\le \kappa\left[\frac{1/x(t)^a - 1/x^{*a}}{x(t)-x^*} + \frac{x(t)^{b+1}c(t)^{-b} - x^{*b+1}c^{-b}}{x(t)-x^*}\right]|x(t)-x^*| \\ &\le -\kappa\left[\frac{x^{*-a} - x(t)^{-a}}{x(t)-x^*} - \frac{x(t)^{b+1}c(t)^{-b} - x^{*b+1}c^{-b}}{x(t)-x^*}\right]|x(t)-x^*|\end{aligned}$$

Therefore, according to the results of Theorem 1, the system is globally asymptotically stable.

## IV. NUMERICAL ILLUSTRATIONS

Consider a simple network topology with a single source that uses a single link. The utility function is of the form $U(x) = -1/(ax^a)$ with $a$=1.5. The price function is of the form $p(x) = (x/c(t))^b$ with $b$=0.8. In addition, we assume that $\kappa$=1, and the delays are given by $\tau = 3\,s$ and $T = 2$ s. Consider that the link capacity is set to be $c(t) = 5 - x(t)$ which is an adaptive virtual queue (AVQ) controller [29]. The initial condition is selected as $x_0(\theta) = 1$ , for $\theta\in[-3,0]$ and $c_0(s) = 4$ , for $s\in[-2,0]$ .Therefore, we have $x^* = 1.1059$ , and $c^* = 3.8941$ .Fig. 1 shows the simulation results for the case that the conditions of Theorem 2 is not satisfied. It is obvious that the system is not stable and there exist oscillations in the flow rates. If we change the parameter $b$ to 0.2, the condition of Theorem 2 in the range of $x$ and $c$ will be satisfied, and the system will be asymptotically stable as shown in Fig. 2.

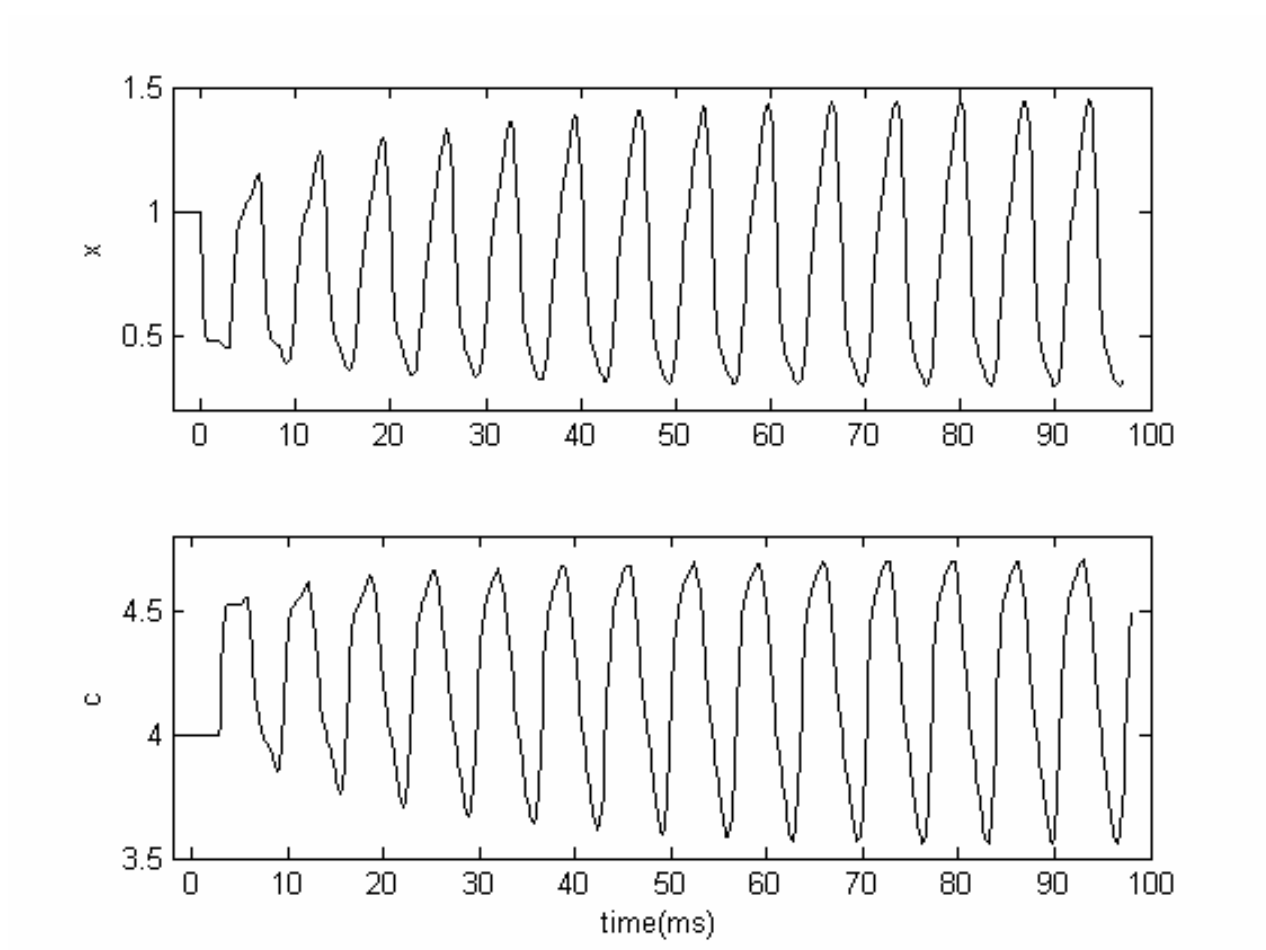

Figure 1. The source flow rate and the capacity of the link for an unstable system

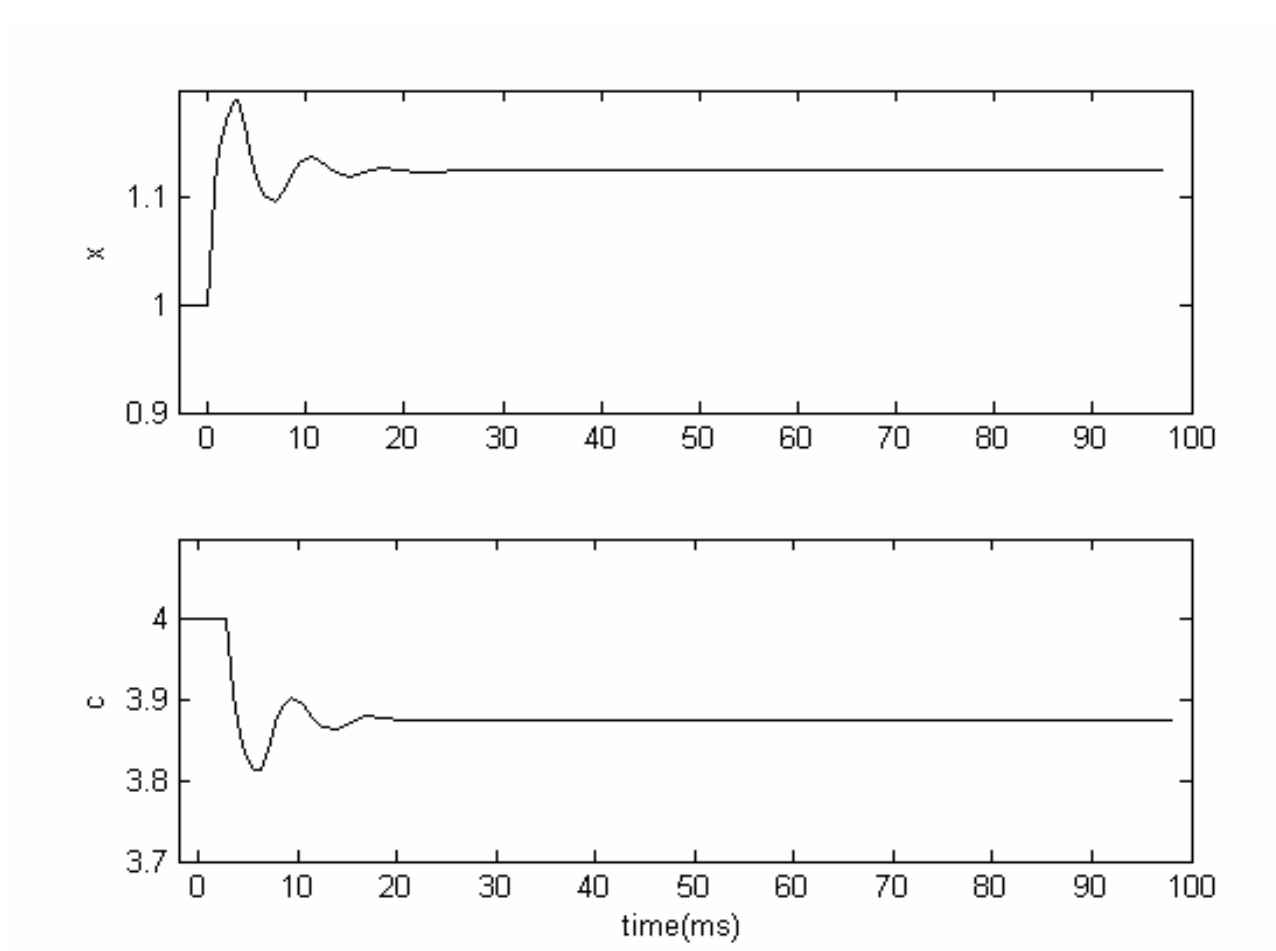

Figure 2. The source flow rate and the capacity of the link for a stable system

## V. CONCLUSION

In this paper, the global stability conditions have been obtained for the congestion control system as a class of nonlinear time-delay systems. The model consists of a time-varying capacity of the link and a fixed communication delay. Using the Lyapunov-Krasovskii's stability theorem, we have shown that the system is globally asymptotically stable when some delay-independent conditions on the source and link parameters are satisfied. The results can be extended to a general multi-source multi-link network.

## REFERENCES

[1] S. Shakkottai, and R. Srikant, “How good are deterministic fluid models of internet congestion control,” in Proc. IEEE INFOCOM, New York, NY, 2001.

[2] P. Ranjan, E. H. Abed, and R. J. La, “Nonlinear instabilities in TCP-RED,” IEEE/ACM Trans. Networking, vol. 12, no. 6, pp. 1079-1092, Dec. 2004.

[3] F. Kelly, A. Maulloo, and D. Tan, “Rate control in communication networks: shadow prices, proportional fairness and stability,” J. Oper. Res. Soc., vol. 49, pp. 237-252, May 1998.

[4] F. Kelly, “Models for a self-managed Internet,” Phil. Trans. Royal Soc. A, vol. 358, pp. 2335–2348, 2000.

[5] S.H. Low and D.E. Lapsley, “Optimization flow control—I: Basic algorithm and convergence,” IEEE/ACM Trans. Networking, vol. 7, no. 6, pp. 861–874, Dec. 1999.

[6] F. Paganini, “On the stability of optimization-based flow control,” in Proc. American Control Conf., Arlington, VA, 2001.

[7] R. Johari and D. Tan, “End-to-end congestion control for the internet: Delays and stability,” IEEE/ACM Trans. Networking, vol. 9, pp. 818–832, Dec. 2001.

[8] L. Massoulie, “Stability of distributed congestion control with heterogeneous feedback delays,” IEEE Trans. Automatic Control, vol. 47, pp. 895–902, June 2002.

[9] G. Vinnicombe, “On the stability of end-to-end congestion control for the Internet,” Univ. Cambridge, Cambridge, U.K., Tech. Rep. CUED/F-INFENG/TR.398, 2000.

[10] G. Vinnicombe, “On the stability of network operating TCP-like congestion control,” in Proc. IFAC World Congress on Automatic Control, Barcelona, Spain, 2002.

[11] T. Alpcan and T. Basar, “A utility-based congestion control scheme for internet-style networks with delay,” in *Proc. IEEE INFOCOM*, San Francisco, CA, pp. 1039-1048, Mar. 2003.

[12] S. Deb, and R. Srikant, “Global stability of congestion controllers for the internet,” *IEEE Trans. Automatic Control*, vol. 48, no. 6, pp. 1055-1060, June 2003.

[13] X. Fan, M. Arcak, and J. T. Wen , “*Lp* stability and delay robustness of network flow control,” in *Proc. IEEE CDC*, Maui, Hawaii, pp. 3683-3688, Dec. 2003.

[14] C. V. Hollot and Y. Chait, “Nonlinear stability analysis for a class of TCP/AQM networks,” in *Proc. IEEE CDC*, Orlando, FL, pp. 2309-2314, Dec. 2001.

[15] F. Mazenc and S.-I. Niculescu, “Remarks on the stability of a class of TCP-like congestion control models,” in *Proc. IEEE CDC*, Maui, Hawaii, pp. 5591-5594, Dec. 2003.

[16] A. Papachristodoulou, “Global stability analysis of a TCP/AQM protocol for arbitrary networks with delay,” in *Proc. IEEE CDC*, Paradise Island, Bahamas , Dec. 2004.

[17] A. Papachristodoulou, J.C. Doyle, S.H. Low, “Analysis of nonlinear delay differential equation models of TCP/AQM protocols using sums of squares,” in *Proc. IEEE CDC*, Paradise Island, Bahamas, pp. 4684-4689, Dec. 2004.

[18] M. Peet, and S. Lall, “Global stability analysis of a nonlinear model of Internet congestion control with delay,” *IEEE Trans. Automatic Control*, vol. 52, no. 3, pp. 553-559, Mar. 2007.

[19] P. Ranjan, R.J. La, and E.H. Abed, “Global stability conditions for rate control with arbitrary communication delay,” *IEEE/ACM Trans. Networking,* vol. 14, no. 1, pp. 94-107, Feb. 2006.

[20] Z. Wang and F. Paganini, “Global stability with time-delay in network congestion control,” in *Proc. IEEE CDC,* pp. 3632-3637, Dec. 2002.

[21] Z. Wang and F. Paganini, “Global stability with time-delay of a primal-dual congestion control,” in *Proc. IEEE CDC,* pp. 3671-3676, 2003.

[22] Z. Wang, and F. Paganini, “Boundedness and Global Stability of a Nonlinear Congestion Control with Delays,” *IEEE Trans. Automatic Control,* vol. 51, no. 9, pp. 1514-1519, Sept. 2006.

[23] J. Wen, and M. Arcak, “A unifying passivity framework for network flow control,” in *Proc. IEEE INFOCOM*, San Francisco, CA, pp. 1156-1166, 2003.

[24] L. Ying, G.E. Dullerud, and R. Srikant, “Global stability of internet congestion controllers with heterogeneous delays,” in *Proc. American Control Conf.*, Boston, Massachusetts, pp. 2648-2953, 2004.

[25] L. Ying, G.E. Dullerud, and R. Sirkant, “Global stability of a Internet congestion controllers with heterogeneous delay,” *IEEE/ACM Trans.Networking*, vol. 14, pp. 579-591, June 2006.

[26] M. Peet, and S. Lall, "Constructing Lyapunov functions for nonlinear delay-differential equations using semidefinite programming," in Proc. IFAC NOLCOS, 2004.

[27] J. K. Hale, and S. M.Verduyn Lunel, Introduction to functional differential equations. New York, NY, Springer Verlag, 1993.

[28] Kunniyur, S., Srikant, R., "Analysis and design of an adaptive virtual queue algorithm for active queue management," in Proc. ACM SIGCOMM, San Diego, CA, pp. 123–134, 2001.